\documentclass[aps,twocolumn]{revtex4}
\usepackage{graphicx}
\usepackage[latin1]{inputenc}
\begin{document}

\title{Spectral Analysis and the Dynamic Response of Complex Networks}

\author{M.A.M. de Aguiar$^{1,2}$ and Y. Bar-Yam$^1$}

\affiliation{$^1$ New England Complex Systems Institute,
Cambridge, Massachusetts 02138\\ $^2$Instituto de F\'isica Gleb
Wataghin, Universidade Estadual de Campinas, 13083-970 Campinas,
S\~ao Paulo, Brazil}

\begin{abstract}

The eigenvalues and eigenvectors of the connectivity matrix of
complex networks contain information about its topology and its
collective behavior. In particular, the spectral density
$\rho(\lambda)$ of this matrix reveals important network
characteristics: random networks follow Wigner's semicircular law
whereas scale-free networks exhibit a triangular distribution. In
this paper we show that the spectral density of hierarchical
networks follow a very different pattern, which can be used as a
fingerprint of modularity. Of particular importance is the value
$\rho(0)$, related to the homeostatic response of the network: it
is maximum for random and scale free networks but very small for
hierarchical modular networks. It is also large for an actual
biological protein-protein interaction network, demonstrating that
the current leading model for such networks is not adequate.

\end{abstract}

\maketitle


The network concept has been gaining recognition as a fundamental
tool in both biological and social sciences, where the theory of
complex systems finds fertile ground. Biological examples include
food webs in ecology \cite{fw}, nervous systems \cite{koch},
cellular metabolism \cite{jeong00}, protein conformation
\cite{scala01} and a protein-protein interaction network
\cite{bionet}. Social networks include scientific collaboration,
citation, problem solving and linguistic networks \cite{scn}. Most
biological and social networks studied are not randomly connected,
they follow a {\it scale free} behavior (see \cite{bararev} and
references therein). In random networks the probability that a
node has $k$ connections, $P(k)$, is Poisson distributed and,
therefore, every node has about the same number of connections. In
scale free networks $P(k)$ follows a power law, a property that
can be constructed by sequential {\it preferential attachment} of
nodes, where new nodes are more likely to connect to already
highly connected ones. The properties of such networks are often
characterized by the presence of a few highly connected nodes, the
{\it hubs}, whereas most of the remaining nodes have a small
number of connections. The importance of such networks, originally
couched in terms of robustness of static connectivity to failure
despite sensitivity to attack \cite{albert2000}, may perhaps be
better characterized in terms of their response dynamics, that
provides both robustness {\it and} sensitivity\cite{baryam2004}.

Although scale free networks describe several statistical
properties of biological networks, they fail to take into account
one important aspect, namely, the modularity exhibited by most
complex systems \cite{simon,yaneerbook}. The concept of modularity
assumes that the full network of interactions can be partitioned
into a number of sub-networks or modules. Each module is composed
of several elements which are more interconnected than they are
connected to the rest of the network. In real systems a module is
expected to perform an identifiable task, separable from the
functions of other modules \cite{yaneerbook,hart99}. Modular
systems may be organized in a structural hierarchy, with multiple
levels of modular decomposition. Molecules, organelles, cells,
tissues, organs and organisms, families, communities, etc., are an
example of such a hierarchy of structures. Networks incorporating
both modular hierarchy and scale free character were recently
discussed by Barab\'asi \cite{bara} (see also \cite{bara04}). One
property often used to characterize modular networks is their
clustering coefficient---the degree to which neighbors of a node
are connected to each other---which is larger than that of generic
scale-free models.

In this work we investigate the spectral properties of modular
networks. We show that the density of states of the connectivity
matrix (particularly its randomized version where elements are set
to $\pm 1$) provides a connection between the structure and the
dynamic response of a network. This enables us to distinguish
between various models and actual systems in a manner that may be
directly relevant to considering the behavior of system response
to perturbations. In particular, we are able to distinguish
clearly between random, scale-free and modular networks. However,
none of these standard model networks capture the properties of an
actual protein-protein interaction network.


The connectivity (or adjacency) matrix $A$ represents the topology
of the system, indicating which variables are interconnected. It
is defined as $A_{ij}=1$ if nodes $i$ and $j$ are connected and
zero otherwise. The spectral properties of this network may be
used to characterize the topology. If we consider the network as
an influence network, where each link may have a strength and
phase that is not specified, a model of the interactions between
nodes $A_R$ can be constructed from $A$ by changing each of the
entries $1$ of $A$ into $-1$ with $50\%$ probability (keeping
$A_{ij}=A_{ji}$, since they represent the same connection). The
spectral properties of $A_R$ contain information about the
dynamics of the network. If the network is in equilibrium and a
perturbation is introduced, this perturbation propagates through
the nodes according to $A_R$. In a linear approximation the state
of the nodes are updated according to $x_i^{t+1} = \sum_j A_{Rij}
x_j^t$. Below we study the spectral properties of $A$ and $A_R$
and show they are in many cases similar, or otherwise can be
related.

The smoothed density of states of the network is defined by
\begin{equation}
\bar{\rho}_{\epsilon}(\bar{\lambda}) = \frac{1}{N} \, \sum_i
\delta_{\epsilon}(\bar{\lambda}-\lambda_i)
\end{equation}
where $\lambda_i$ are the eigenvalues of the connectivity matrix
and $N$ is the total number of nodes. Since the $A$ is symmetric
all eigenvalues are real. $\delta_{\epsilon}(x)$ is a smoothed
delta function that tends to the real Dirac delta as $\epsilon
\rightarrow 0$. Choosing $\epsilon$ to be a few units of the mean
level spacing produces a smooth level density even for small
networks, which is easier to visualize than the spiked density
produced by the delta functions. Following Farkas et al
\cite{farkas2001} we define scaled variables $\rho$ and $\lambda$
by
\begin{equation}
\lambda = \displaystyle{\frac{\bar{\lambda}}{\sqrt{Np(1-p)}}}
\qquad \rho = \bar{\rho} \sqrt{Np(1-p)}
\end{equation}
where $p=\bar{k}/N$ is the average number of links per node
divided by the total number of nodes. For random networks the
density of states can be computed analytically from random matrix
theory and the result is the so called Wigner's semicircular law.
In the scaled variables it becomes simply $\rho(\lambda) =
\sqrt{4-\lambda^2}/2\pi$ if $|\lambda| < 2$ and zero otherwise.


Figure 1 shows the density of states for four different networks.
All networks have $N = 1024$ nodes, except for the protein-protein
network which has $N=1297$. Fig.1(a) shows $\rho(\lambda)$ for a
random network with $p=0.0057$, following closely Wigner's
semicircular law. Fig.1(b) shows a Scale Free network with
$p=0.0058$, exhibiting a triangular profile \cite{farkas2001}.
Fig.1(c), corresponding to Barabasi's hierarchical network
\cite{bara}, has a peculiar density, that we shall discuss in more
detail. Finally fig.1(d) shows $\rho(\lambda)$ for a
protein-protein interaction network \cite{bionet} and also has a
distinct behavior, looking more like a superposition of two
independent scale free networks.
\begin{figure}
\includegraphics[width=3cm,angle=-90]{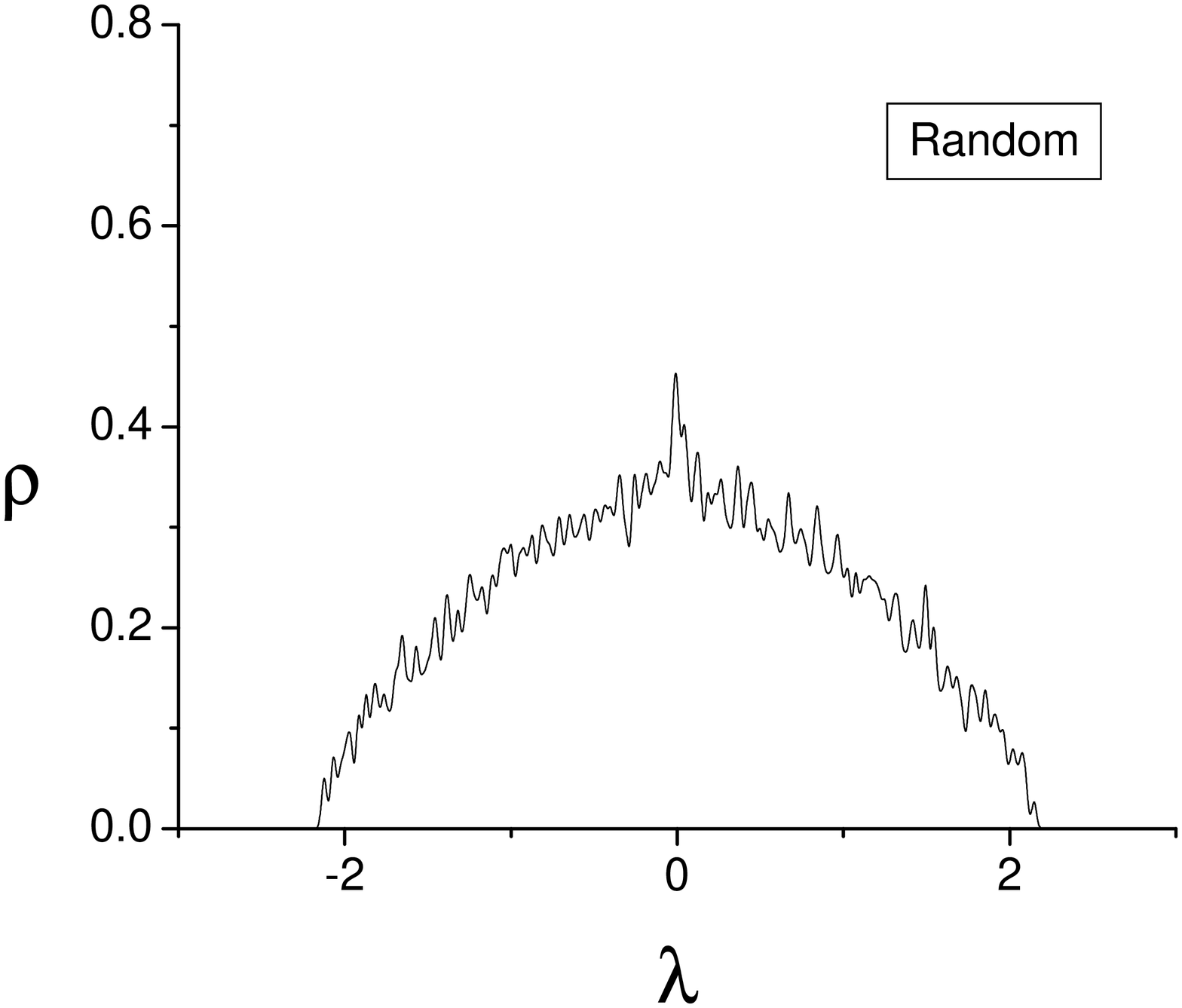}
\includegraphics[width=3cm,angle=-90]{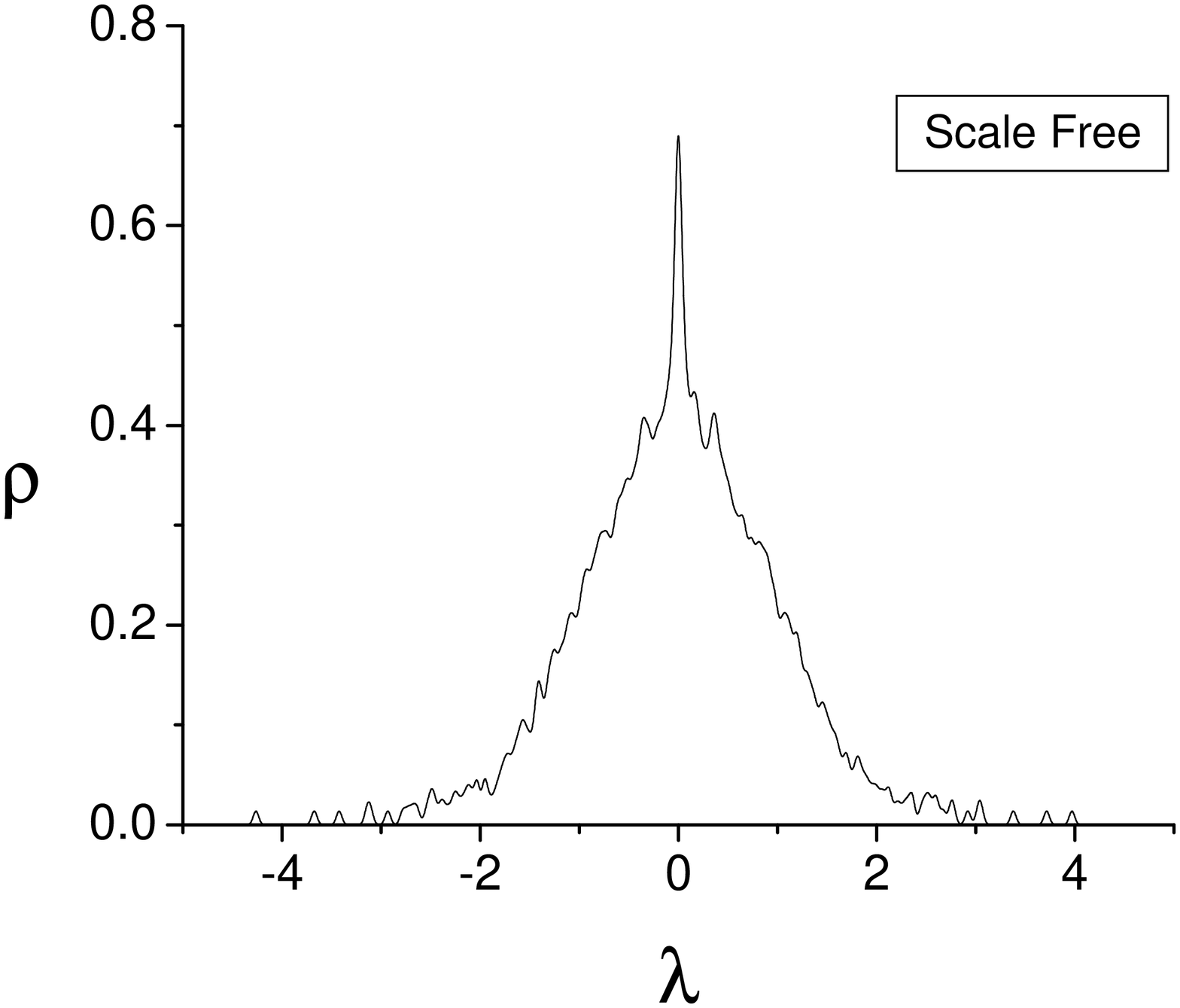}
\includegraphics[width=3cm,angle=-90]{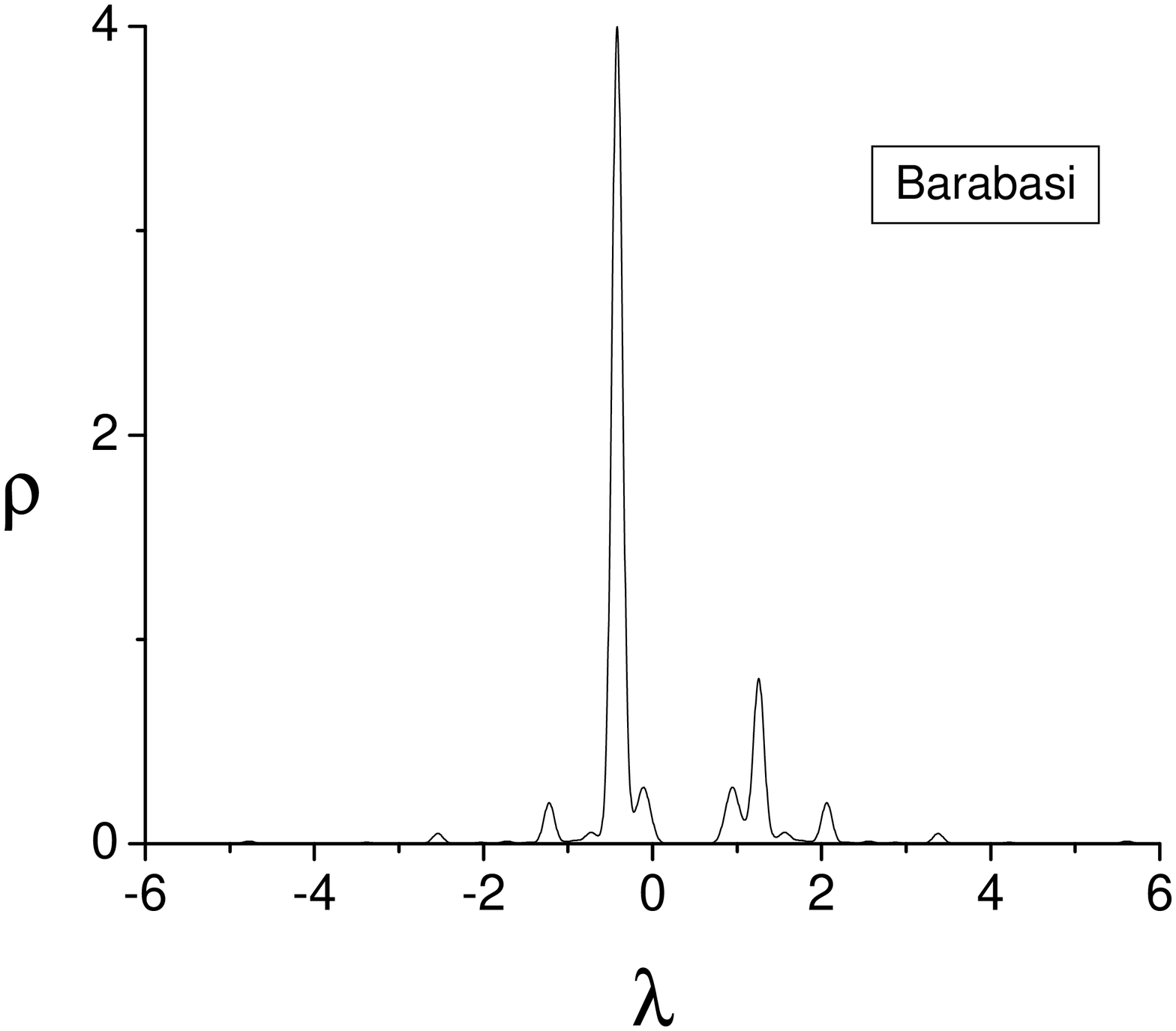}
\includegraphics[width=3cm,angle=-90]{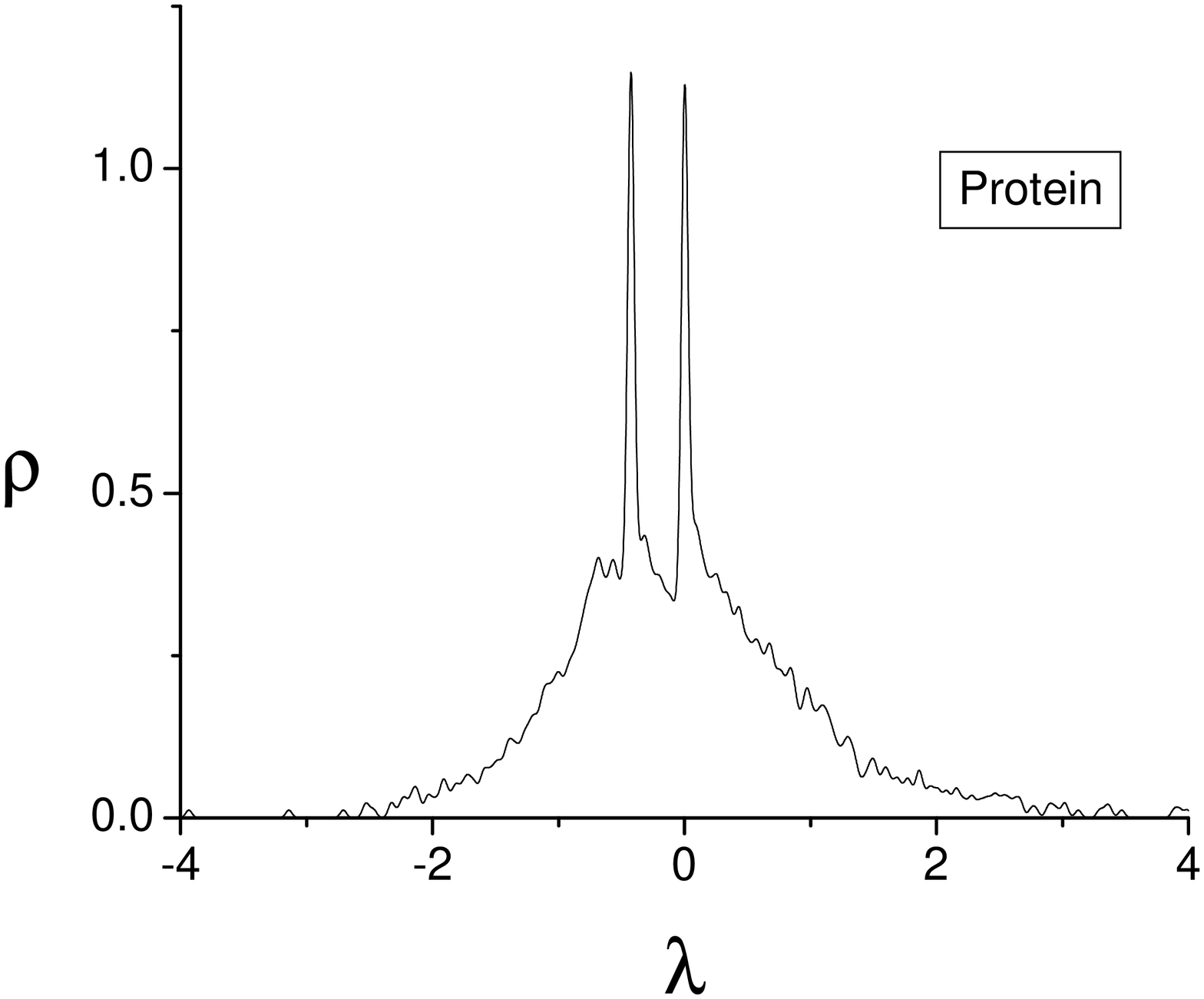}
\label{fig1} \caption{Smoothed density of state for a random,
scale free, Barabasi's hierarchical network (all with $1024$
nodes) and the protein-protein interaction network (with $1297$
nodes).}
\end{figure}

Figure 2 shows the density of states for the same networks
obtained with the randomized connectivity matrices $A_R$. For each
network we diagonalized 20 matrices with random distributions of
$\pm 1$'s and calculated the average density over this ensemble.
The averaged density satisfies $\rho(\lambda)=\rho(-\lambda)$. The
scale free and random networks are not sensitive to sign
randomization, since their original spectra are already symmetric.
\begin{figure}
\includegraphics[width=3cm,angle=-90]{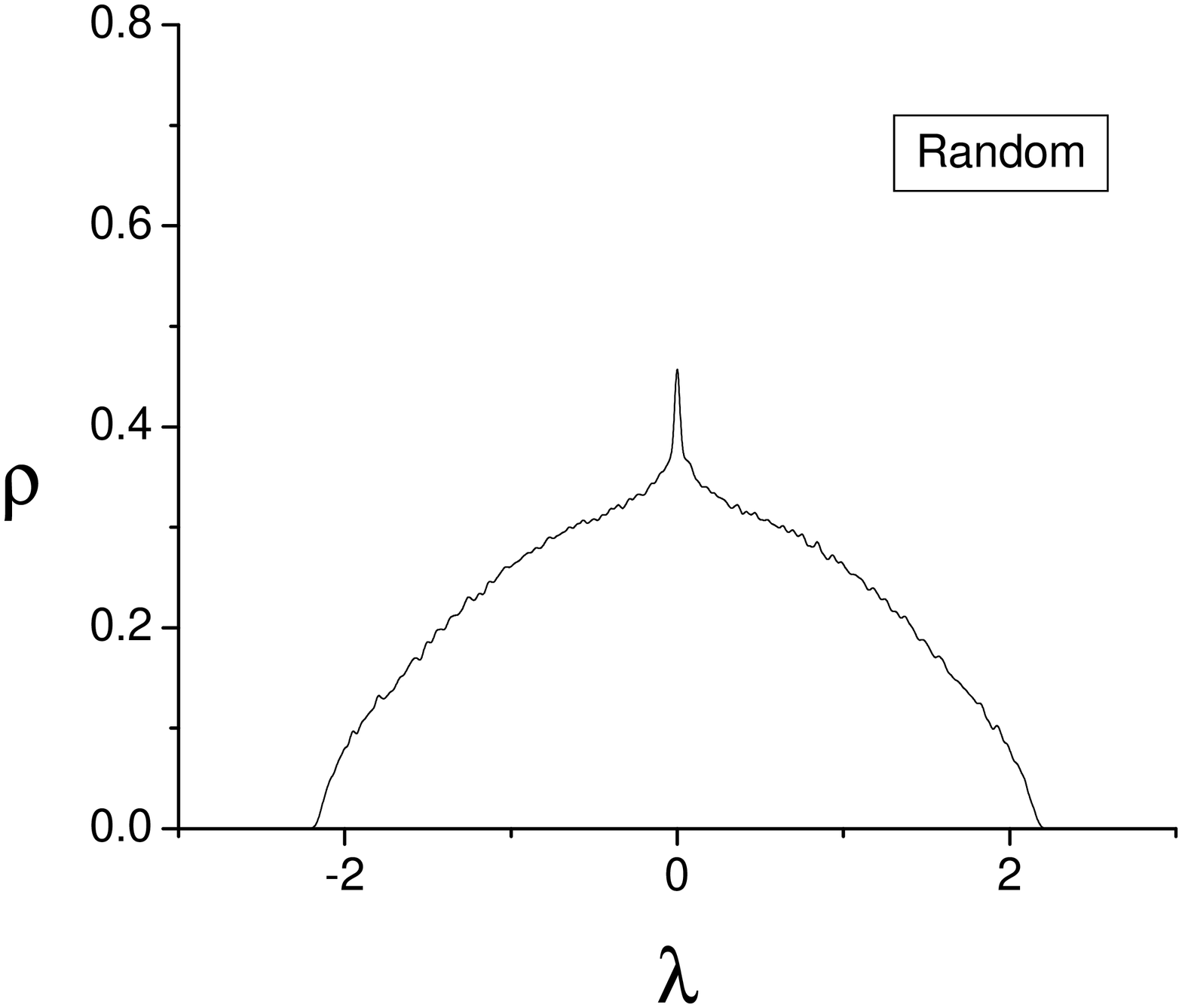}
\includegraphics[width=3cm,angle=-90]{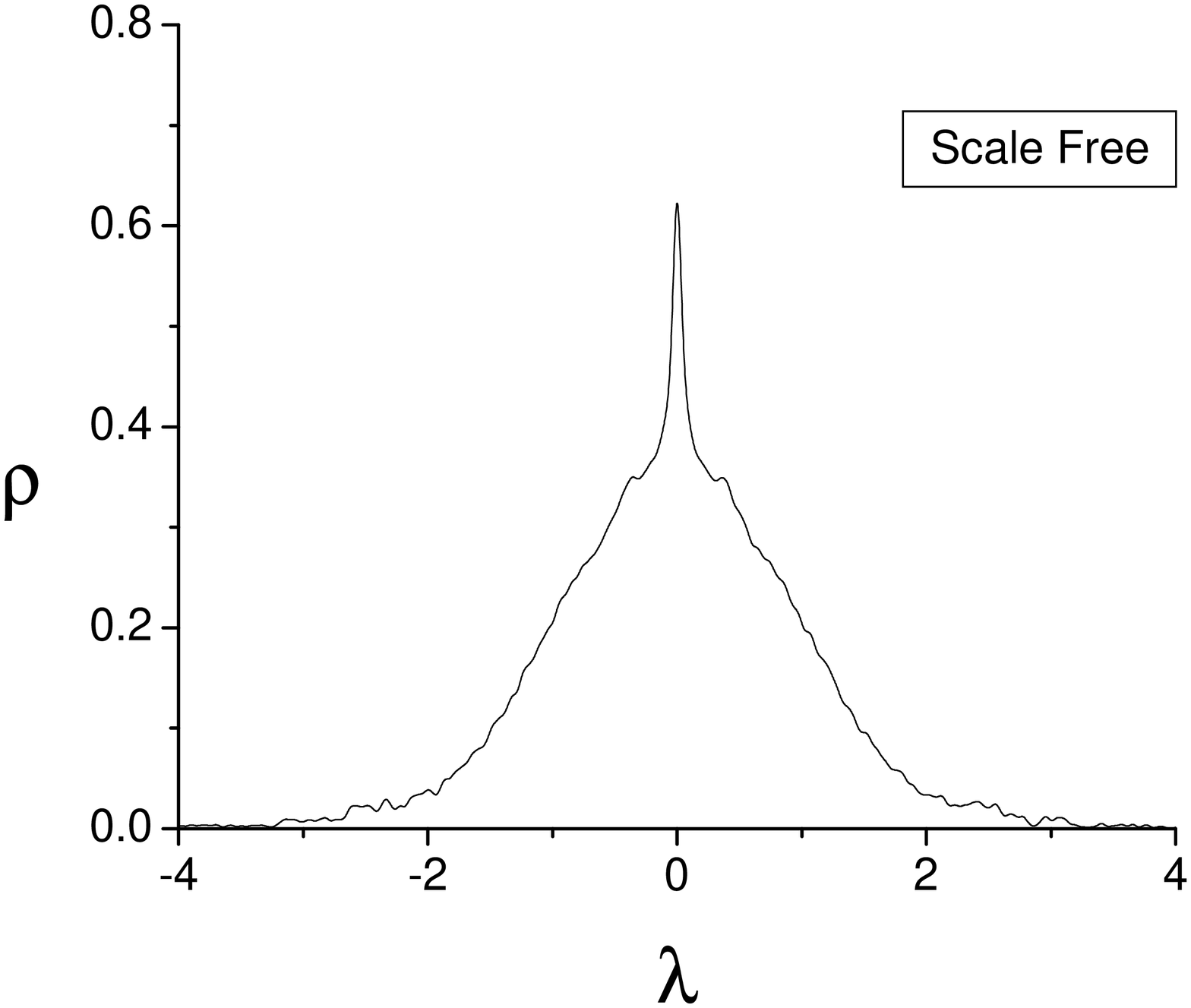}
\includegraphics[width=3cm,angle=-90]{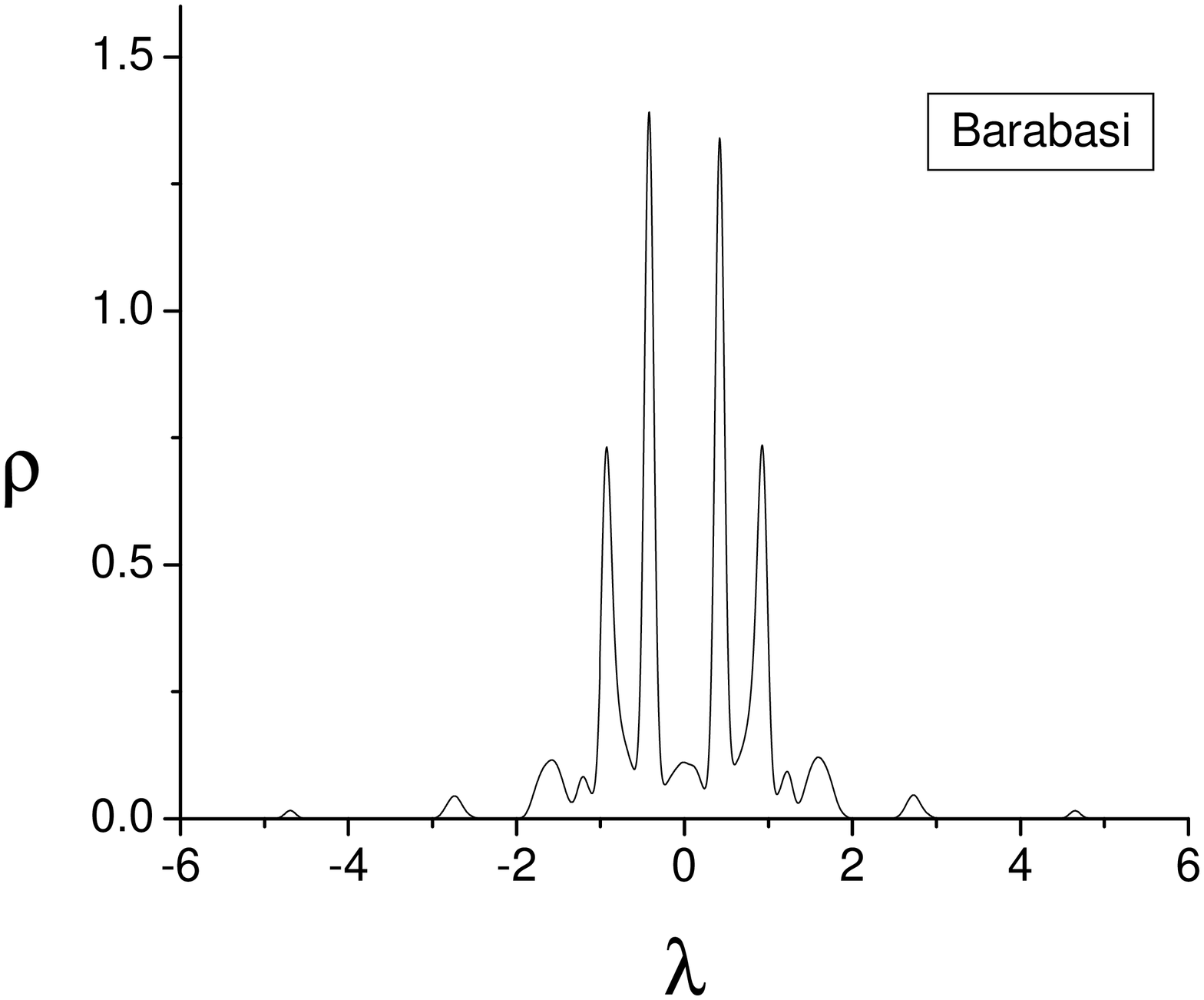}
\includegraphics[width=3cm,angle=-90]{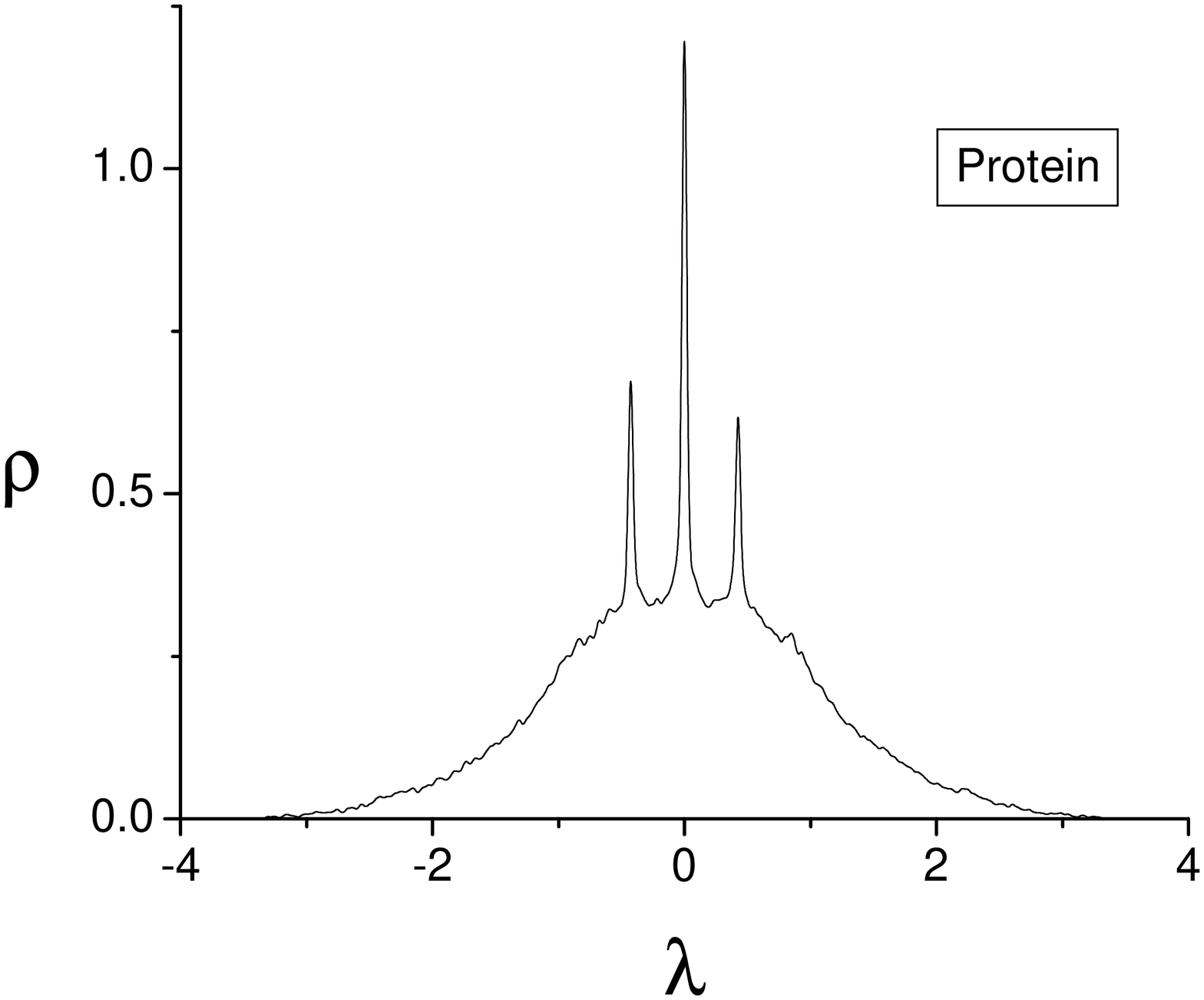}
\label{fig2} \caption{Smoothed density of state for the {\it
randomized} networks of Fig.1.}
\end{figure}
Barabasi's hierarchical network density of states, on the other
hand, changes considerably. It keeps the minimum at $\lambda=0$,
whereas all other networks have a peak there. Also, the density
has sharp peaks with high intensity at certain values of
$|\lambda|$, becoming very small away from the peaks. The
biological network also has an interesting structure, deviating
from the pure scale free case. However, in contrast to Barabasi's
network it has a peak at $\lambda=0$.


Barabasi's hierarchical network is built from a fully connected
network with 4 nodes. This unit is then replicated three times and
the four identical networks are connected together. The network
thus formed is then viewed as the new unit, and the replicating
and connecting process is repeated \cite{bara}. Although the exact
repetition of this process is artificial, one expects real modular
networks to exhibit some type of self-similar structure. In what
follows we shall show that networks built from such basic units
have indeed a very characteristic spectrum, that can be used to
identify its modular nature.


Consider first a fully connected network with $N$ nodes. The
connectivity matrix is $(A_{N})_{ij}=1-\delta_{ij}$. The
eigenvalues of $A_N$ can be calculated immediately and we find
$\lambda_1=N-1$, $\lambda_2=\lambda_3=...=\lambda_N=-1$. The first
eigenvector $|w_1\rangle$, corresponding to the largest eigenvalue
$\lambda_1$, has components $w_{1,i}=1$. All the other
eigenvectors are degenerate and satisfy $\sum_i w_{j,i}=0$. It is
possible to choose them so as to have very few non-zero elements.
The linear update equations $x^{t+1} = A_N x^t$ decouples into
$y_i^{t+1} = \lambda_i y_i^t$ and $y_i^t = \lambda_i^t y_i^0$. The
dominant mode is the `center of mass' $y_1$, meaning that the
network synchronizes and responds as a unit to the perturbation.
All other modes involve fewer nodes and correspond to oscillations
of fixed amplitude. The density of states for a fully connected
network has only two peaks: one at $\lambda=-1$ and the other at
$\lambda=N-1$, the former being $N-1$ times larger than the
latter.
\begin{figure}
\includegraphics[width=3cm,angle=-90]{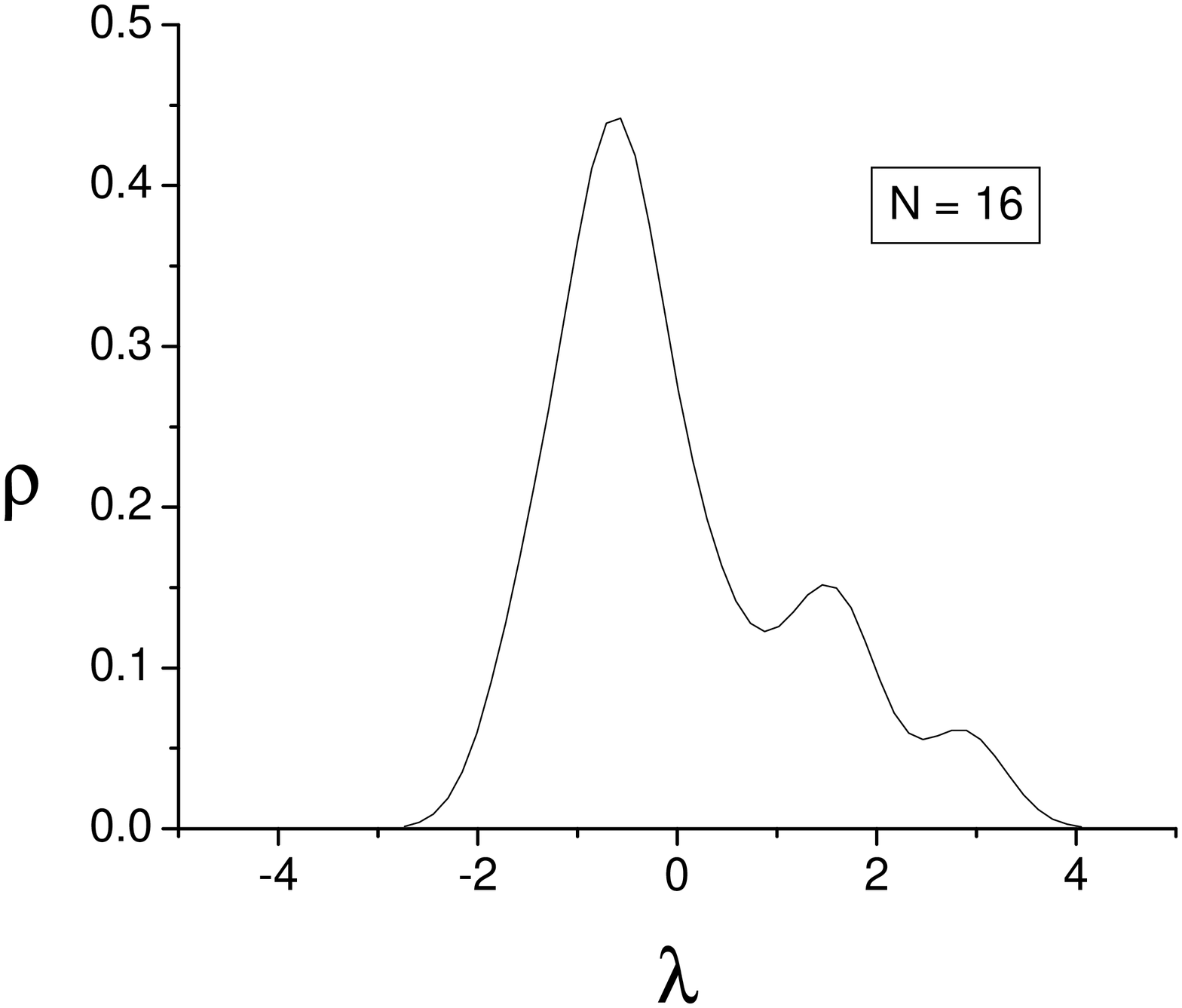}
\includegraphics[width=3cm,angle=-90]{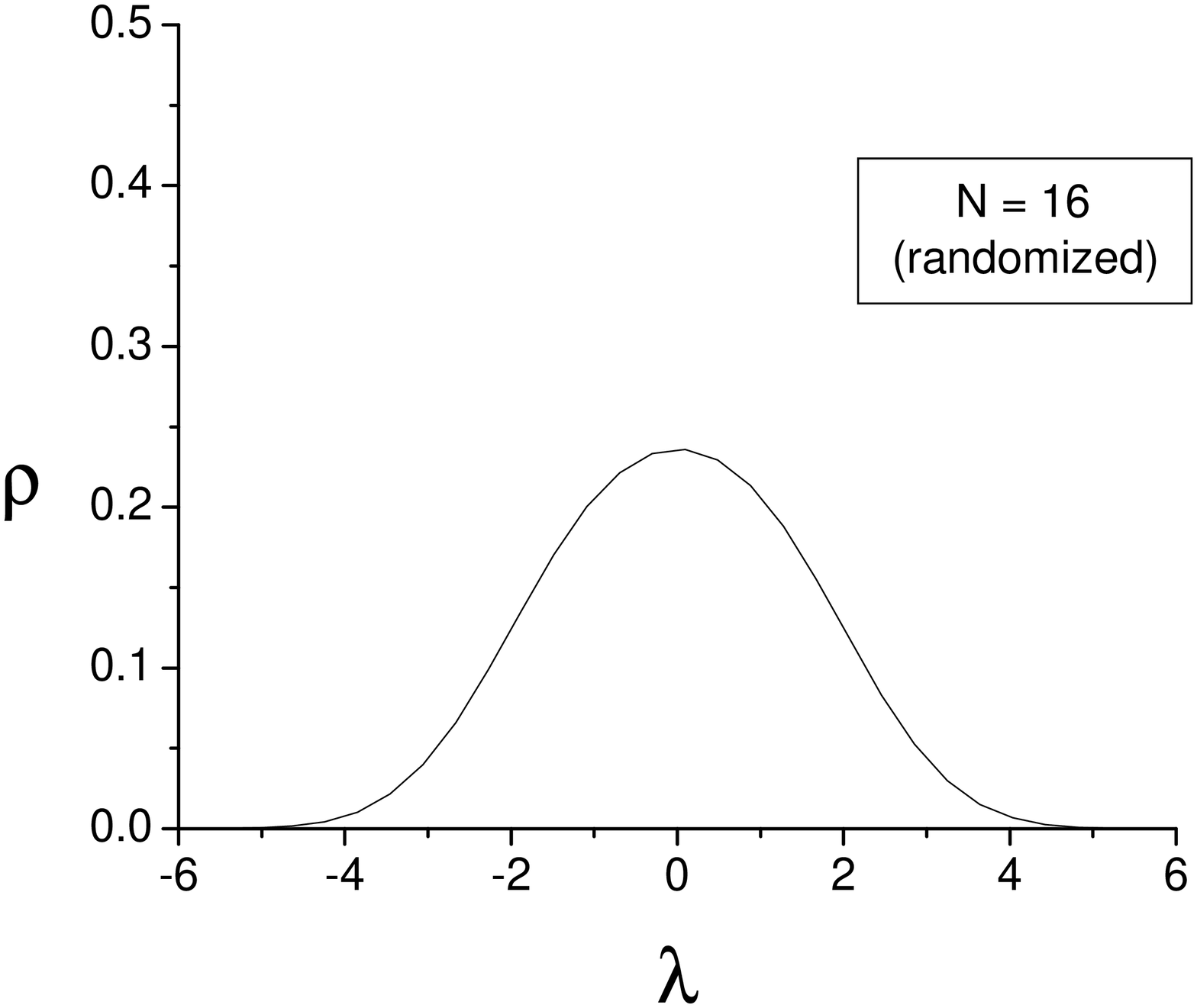}
\includegraphics[width=3cm,angle=-90]{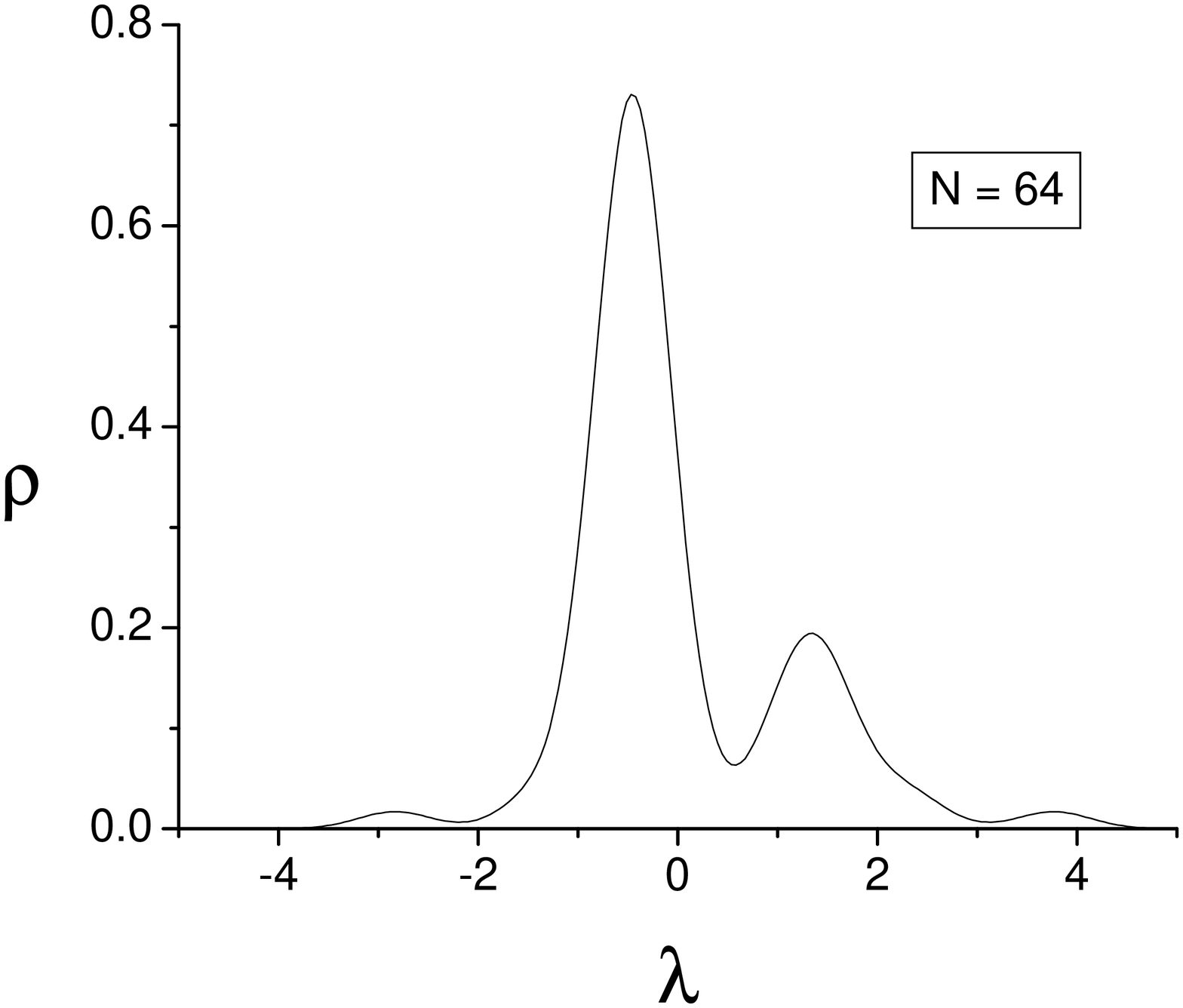}
\includegraphics[width=3cm,angle=-90]{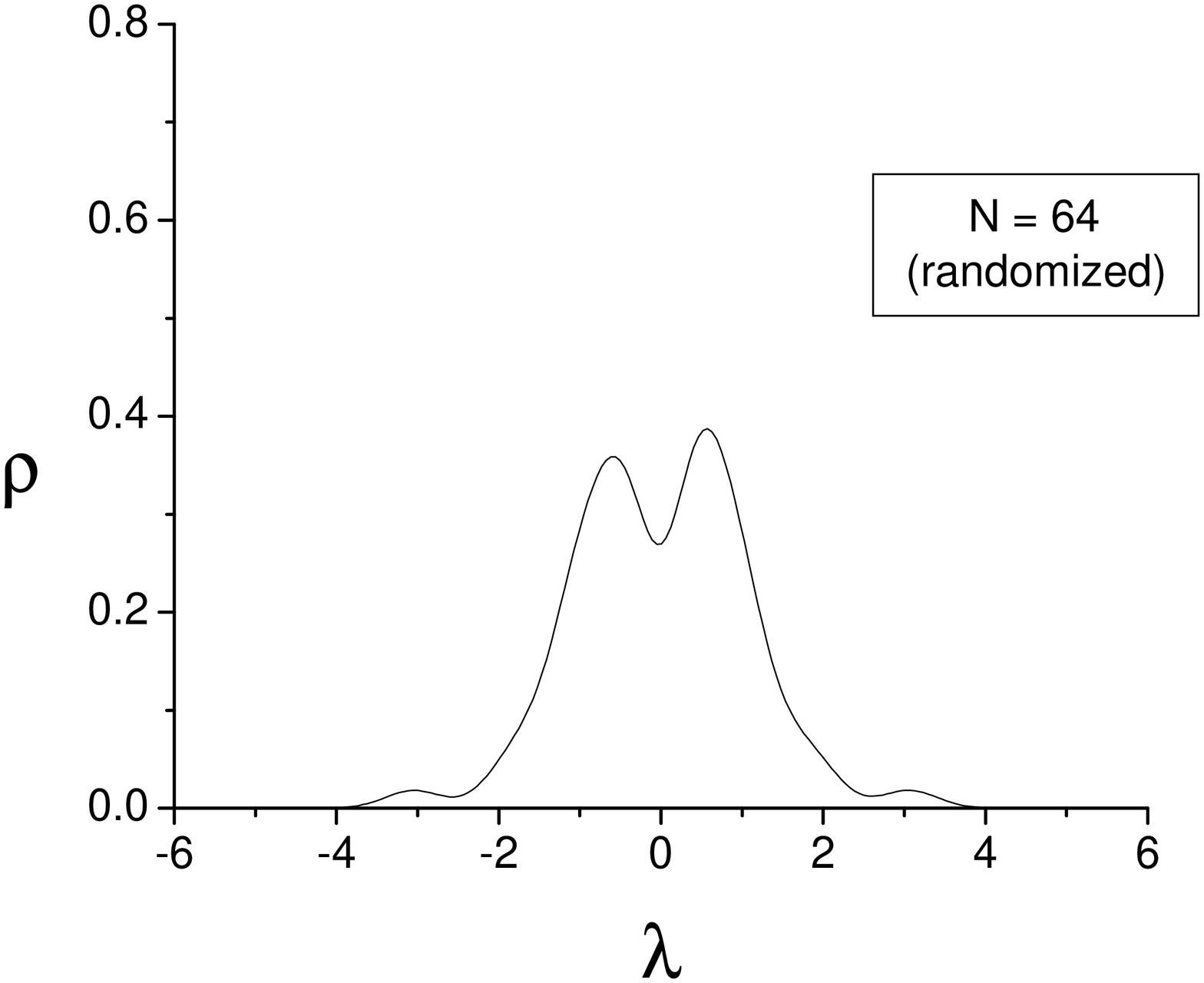}
\includegraphics[width=3cm,angle=-90]{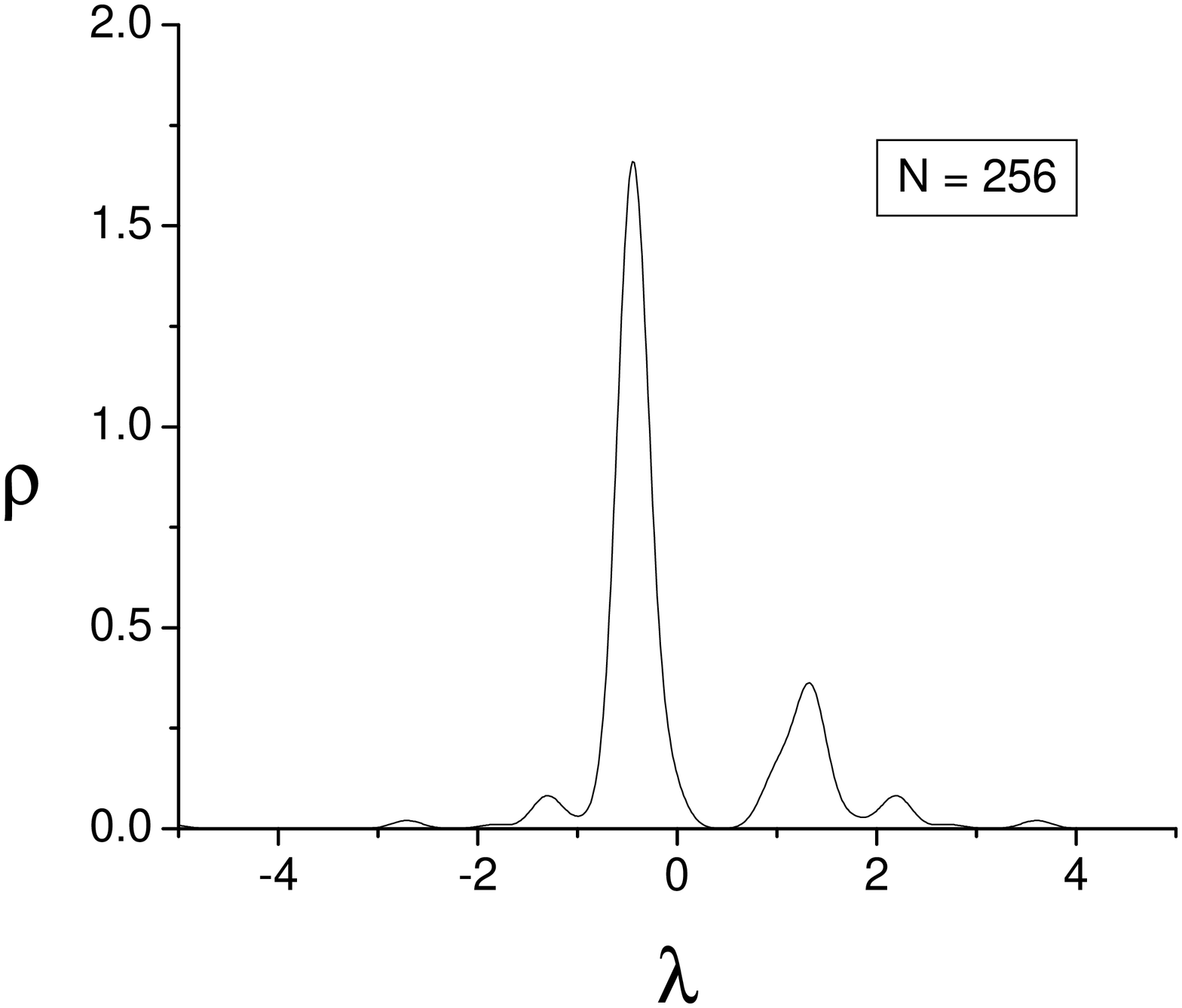}
\includegraphics[width=3cm,angle=-90]{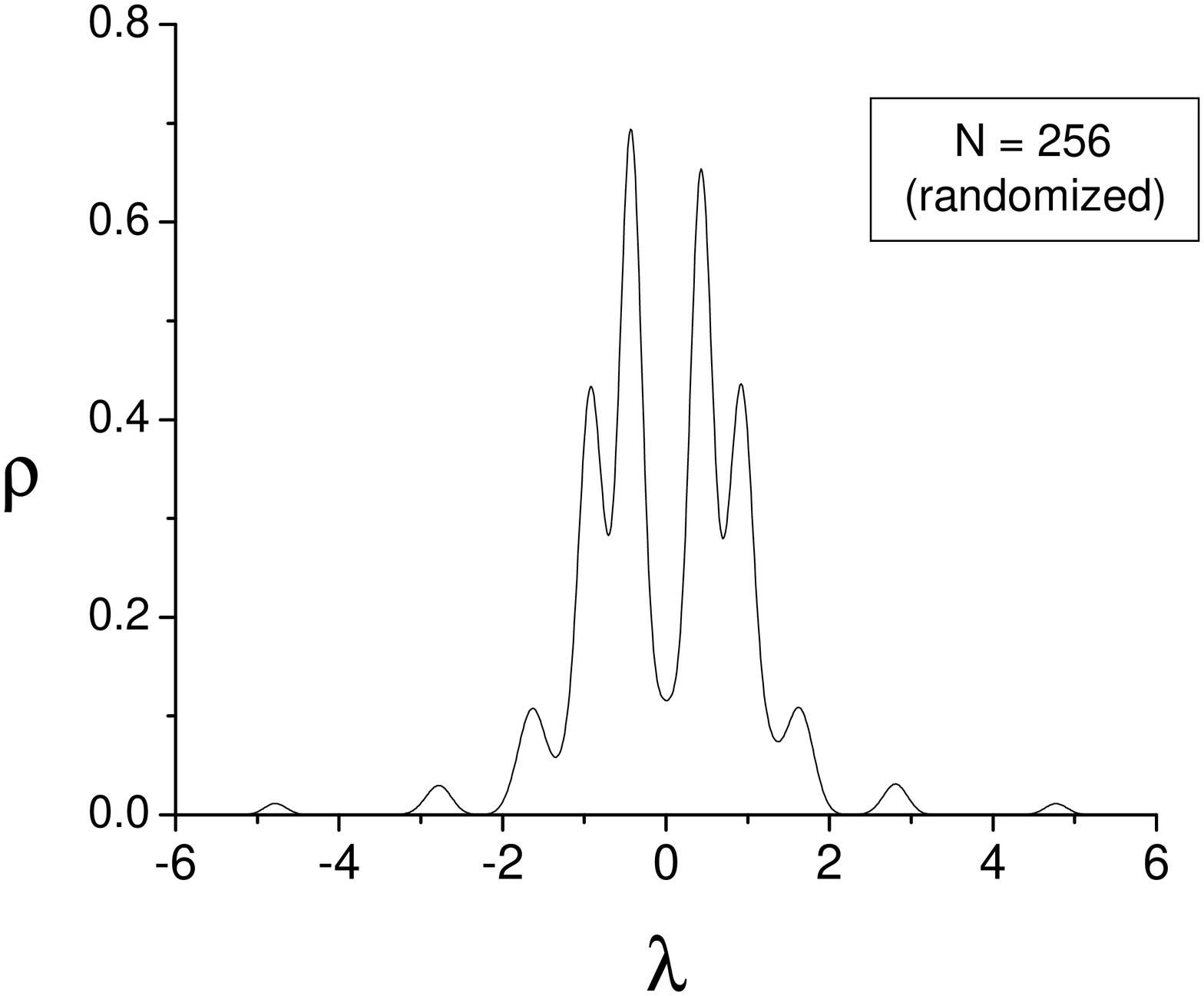}
\label{fig3} \caption{Smoothed density of state for Barabasi's
hierarchical network for the original and the randomized
connectivity matrix with $N=16$, $64$ and $256$ nodes. The
smoothing parameter is 5 times the average level spacing, except
for $N=16$, where it is 2 times the average level spacing.}
\end{figure}


Next we consider {\it star networks}, which are characterized by a
single hub, i.e., a single central node to which all other nodes
are connected. Star networks emerge in systems in which
preferential attachment is superlinear, meaning that the
probability that a new node attaches to old nodes increases faster
than expected by linear preferential attachment \cite{krap}.
Star-like clusters are very common in biological networks (see for
instance \cite{valv}) and their eigenvalues and eigenvectors can
also be computed exactly. In the idealized star network the nodes
connect only to the central node, which we label 1. The
connectivity matrix is given by $A_{i1}=A_{1i}=1$ for
$i=2,3,...,N$ and $A_{ij}=0$ otherwise. The eigenvalues are
$\lambda_1=\sqrt{N-1}$, $\lambda_2=\lambda_3= ...=\lambda_{N-1} =
0$ and $\lambda_N=-\sqrt{N-1}$. The first eigenvector
$|w_1\rangle$ has components $w_{11}=\sqrt{N-1}$ and $w_{1i}=1$
for $i\geq 2$. The last eigenvector $|w_N\rangle$ is given by
$w_{N1}=\sqrt{N-1}$ and $w_{Ni}=-1$ for $i\geq 2$. All the others
degenerate eigenvectors satisfy $w_{j,1}=0$ and $\sum_{i=2}^{N}
w_{j,i} = 0$.

Now we consider a network whose connectivity matrix has a modular
organization consisting of 4 main blocks, each one very similar to
the others. The number 4 is chosen only for comparison with
Barabasi's model, but could be any number. We assume that
the blocks are fully connected, so that we know their eigenvectors
and eigenvalues when they are decoupled. Let
$|w_i^{\alpha}\rangle$ be the $i-th$ eigenvector of the block
labelled by $\alpha$. Since the blocks are all identical, the
eigenvalues are degenerate: $\lambda_1^{\alpha} = M-1$ and
$\lambda_i^{\alpha} = -1$ for $i \neq 1$, where $M$ is the
dimension of the blocks. The connectivity matrix can represented
in block form by
\begin{equation}
\label{mat4}
A = \left(
\begin{array}{cccc}
A_M &  v_{12}  &  v_{13}  & v_{14} \\
v^T_{12}   & A_M &  v_{23}  & v_{24} \\
v^T_{13}   &  v^T_{23}  & A_M & v_{34} \\
v^T_{14}   &  v^T_{24}  &  v^T_{34}  & A_M
\end{array}
\right) \equiv A^0 + V
\end{equation}
where $A_M$ are fully connected $M$ by $M$ matrices, $A^0$ is the
{\it unperturbed} matrix, with the 4 uncoupled $A_M$ blocks, and
$V$ is a sparse perturbation, representing the weak connection
between nodes of different blocks.

The perturbation breaks the degeneracy between the blocks. The
first eigenvalue becomes $\lambda = \lambda_0 + \mu$ and the
corresponding eigenvector
%
$|v_1^{\alpha}\rangle = \sum_{\beta} a_{\alpha\beta}
|w_1^{\beta}\rangle + |\xi \rangle$
%
where the sum over $\beta$ runs over the blocks and represents the
linear combination between the originally degenerate vectors and
the last term is the correction due to the perturbation. Writing
the eigenvalue equation for $|v_1^{\alpha}\rangle$ and keeping
only linear terms in the perturbation $V$ leads to the condition
\begin{equation}
\sum_{\beta} a_{\alpha\beta} \left[ \langle w_1^{\alpha}| V
|w_1^{\beta}\rangle - \mu \delta_{\alpha\beta}\right] = 0 \;.
\label{v1}
\end{equation}

For all the other eigenvectors, whose degeneracy is much bigger,
we write
%
$|v_n^{\alpha}\rangle = \sum_{\beta m} a_{\alpha\beta}^{nm}
|w_m^{\beta}\rangle + |\xi \rangle$
%
where the sum now runs over $\beta$ and $m$, with $n,m\neq 1$. The
eigenvalue equation for this case is
\begin{equation}
\sum_{\beta m} a_{\alpha\beta}^{nm} \left[ \langle w_n^{\alpha}| V
|w_m^{\beta}\rangle - \mu \delta_{\alpha\beta}\delta_{nm} \right]
= 0 \;.
\end{equation}
However, each matrix element $\langle w_n^{\alpha}| V
|w_m^{\beta}\rangle$ is obtained by adding elements of the matrix
$V$ with coefficients that add up to zero. Since $V$ is sparse, we
expect most of these elements to be zero and, when they are not
zero, there will likely be cancellations. Therefore, the
corrections to the eigenvalues are going to be small, and the
density of states of $A$ should still have a large peak around
$\lambda=-1$.

On the other hand, the elements of $|w_1^{\beta}\rangle$ are all
$1$ inside the $\beta$ block and zero outside:
\begin{equation}
\langle w_1^{\alpha}| V |w_1^{\beta}\rangle = \sum_{k,l}
w_{1,k}^{\alpha} V_{kl} w_{1,l}^{\beta}  \equiv K_{\alpha\beta}
\end{equation}
where $K_{\alpha\beta}$ is the number of $1$'s in the block
$v_{\alpha\beta}$. At this point we have to distinguish between
random and scale free networks:\\

\emph{random coupling - } We can assume that all the coupling
blocks $v_{\alpha\beta}$ are similar, so we write $K_{\alpha\beta}
= a$ where $a$ is the average number of $1$'s in each of the $v$
blocks. The $4\times 4$ matrix to be diagonalized in Eq.(\ref{v1})
is identical to the connectivity matrix of a completely connected
network of 4 nodes. Therefore, the 4 uncoupled eigenvalues $M-1$
unfold into 1 eigenvalue $M-1+3a$ and 3 eigenvalues $M-1-a$. For
random coupling we expect three main peaks in the density of
states: a large peak at $\lambda=-1$, a smaller one at $M-1-a$ and
an even smaller one at
$M-1+3a$. \\

\emph{ scale free coupling - } In this case the blocks are
themselves not connected randomly, they attach preferentially to,
say, the first block. The $4\times 4$ matrix to be diagonalized
has the form
\begin{equation}
\label{matsf4} \left(
\begin{array}{cccc}
0 & a & a & a \\
a & 0 & b & b \\
a & b & 0 & b \\
a & b & b & 0
\end{array}
\right)
\end{equation}
where $a >> b$. In first approximation we neglect $b$ and
the resulting matrix is that of a $4\times 4$ star network.
Therefore, the eigenvalues become: $M-1-\sqrt{3}a$, $M-1$ (doubly
degenerate) and $M-1+\sqrt{3}a$. Together they contribute a single
symmetric peak around $M-1$ with half width $\sqrt{3}a$.
Therefore, for scale free modular matrices we expect only two main
peaks in the density of states: a large one at $\lambda=-1$ and a
smaller one at $\lambda=M-1$.

Figure 3 shows the density of states for Barabasi's hierarchical
network with 16, 64 and 256 nodes. The two peaks structure is clear
and consistent with our analysis of a modular scale free network.  The
protein network shown in Fig.1 is certainly not completely
modular. But it is also not generically scale free either.  The two
peaks at zero and $-1$ (in non-scaled units) suggest the existence of
many star like structures (where the eigenvalue 0 abounds) and many
fully connected modules (where the eigenvalue -1 abounds).

\emph{Randomized connectivity matrices}. A similar analysis can
made for the case of the randomized connectivity matrices. For
example, starting from a single fully connected unit of 4 nodes,
the eigenvalue equation can be seen to be
$\lambda^4-6\lambda^2-2\lambda(a_{23}a_{24}a_{34}+a_{12}a_{24}a_{14}+
a_{12}a_{13}a_{23}+a_{13}a_{14}a_{34})
-2(a_{12}a_{13}a_{24}a_{34}+a_{12}a_{14}a_{23}a_{34}+a_{14}a_
{13}a_{24}a_{23})+3=0$. For random $a_{ij}$'s, the term
multiplying $\lambda$ averages to zero, whereas the constant term
in parenthesis averages to either -1 or +1. The averaged equation
is $ \lambda^4-6\lambda^2+1=0$ or $\lambda^4-6\lambda^2+5=0$. The
result is a spectrum with two pairs of symmetric eigenvalues. When
a modular network is constructed out of these random units, we
obtain a density of state with four symmetric peaks. This can be
seen in Fig.3 for Barabasi's network with 16, 64 and 256 nodes.

To summarize, we have introduced some conventional but powerful
tools into the discussion of networks, namely, linear algebra and
perturbation analysis. We have shown that the density of states
contains crucial information not only about the topology of the
network but also about its response to external perturbations. By
comparing $\rho(\lambda)$ for a random, a scale-free and
Barabasi's hierarchical network, we have shown that it exhibits
clear fingerprints of the networks they represent. More
importantly, we have shown that neither of these model networks
can describe the density of states of a real protein-protein
interaction network, showing that better network models are
necessary to understand biological systems. In particular, the
behavior of $\rho(0)$, which indicates that the real biological
network has a robust homeostatic response, is not reproduced by
Barabasi's hierarchical model. Our analysis also indicates the
presence of several star like and fully connected modules in the
biological network, suggesting that these structures might have to
be incorporated explicitly in more realist models.



\begin{thebibliography}{99}

\bibitem{fw} J. Cohen, F. Briand, C. Newman, {\it Community Food
Webs: Data and Theory} (Springer, Berlin, 1990); N.D. Martinez,
Science {\bf 260} 242 (1993); R.J. Willliams and N. D. Martinez,
Nature {\bf 404} 180 (2000); Jose M. Montoya and Ricard V. Sole,
J. Theor. Biol. {\bf 214} 405 (2002).

\bibitem{koch} C. Koch and G. Laurent, Science {\bf 284}, 79 (1999) 96.

\bibitem{jeong00} H. Jeong, B. Tombor, R. Albert, Z. N. Oltvai, and A.-L.
Barab\'asi, Nature {\bf 407} (2000) 651.

\bibitem{scala01} A. Scala, L. A. N. Amaral, and M. Barth\'el\'emy, Europhys.
Lett. {\bf 55} (2000) 594.

\bibitem{bionet} I. Xenarios et al, Nucleid Acids Research {\bf 29}
(2001) 239.

\bibitem{scn} D.J. de S. Price, Science 149, 510 (1965);
M. E. J. Newman, Proc. Natl. Acad. Sci. U.S.A. 98, 404, (2001);
Phys. Rev. E 64, 016131 (2001); Phys. Rev. E 64, 016132 (2001); Ramon
Ferrer and Ricard V. Sole, Phys. Rev. E {\bf 69} 051915 (2004);
D. Braha and Y. Bar-Yam, Phys. Rev. E 69, 016113 (2004). 

\bibitem{bararev} R. Albert and A.-L. Barab\'asi, Rev. Mod. Phys. {\bf 74}
(2002) 47.

\bibitem{albert2000} R. Albert, H. Jeong and A.-L. Barabasi, Nature {\bf 406}, 378
(2000).

\bibitem{baryam2004} Y. Bar-Yam and I. Epstein, {\bf PNAS} 101, 4341 (2004).

\bibitem{simon} H. A. Simon, The Sciences of the Artificial, 3rd ed. (MIT
Press, Cambridge, 1998), Chap. 8.

\bibitem{yaneerbook} Y. Bar-Yam, {\it Dynamics of Complex Systems}, (Perseus Press,
Cambridge, 1997), Chap. 2.

\bibitem{hart99} L. H. Hartwell, J. J. Hopfield, S. Leibler, A. W. Murray,
Nature {\bf 402} (1999) C47 .

\bibitem{bara} E. Ravasz, A. L. Somera, D. A. Mongru, Z. N. Oltvai,
A.-L. Barab\'asi, Science {\bf 297} (2002) 1551

\bibitem{bara04} A.L. Barabasi, Z.N. Oltvai, Nat. Rev. Genet. 5 (2004)
101; A. Fronczak, P. Fronczak, J.A. Holyst,  Phys. Rev. E {\bf 68}
(2003) 046126; J.D. Noh Phys. Rev. E {\bf 67} (2003) 045103.

\bibitem{farkas2001} I.J. Farkas, I. Der\'enyi, A.-L. Barab\'asi, and T. Vicsek,
Phys. Rev. E {\bf 64}, (2001) 026704.

\bibitem{krap} P. L. Krapivsky, S. Redner, and F. Leyvraz, Phys. Rev. Lett. 85,
4629 (2000).

\bibitem{valv} S. Valverde, R.F. Cancho and R.V. Sol\'e, Europhys. Lett. {\bf
60} 512 (2002).

\end{thebibliography}
\end{document}